\documentclass[twocolumn,superscriptaddress,longbibliography,aps,prl]{revtex4-2}

\usepackage{epsfig}
\usepackage{graphicx}
\usepackage[ansinew]{inputenc}
\usepackage{array}
\usepackage[urlcolor=blue,colorlinks=true,citecolor=blue,linkcolor=blue,pdfstartview={FitH},bookmarks=false]{hyperref}
\usepackage{tabularx}
\usepackage{amsmath}
\newcolumntype{C}[1]{>{\centering\let\newline\\\arraybackslash\hspace{0pt}}m{#1}}

\usepackage{xcolor}
\usepackage{ulem}

\begin{document}
\preprint{APS/123-QED}
\title{Magnetic interactions of 4$f$ electrons in the topological insulator chalcogenide Bi$_{2}$Se$_{3}$}

\author{J. C. Souza}
\email[e-mail: ]{Jean.Souza@weizmann.ac.il}
\altaffiliation[Current address: ]{Department of Condensed Matter Physics, Weizmann Institute of Science, Rehovot, Israel}
\affiliation{Instituto de F\'isica \lq\lq Gleb Wataghin\rq\rq, UNICAMP, 13083-859, Campinas, SP, Brazil}

\author{M. Carlone}
\affiliation{POSMAT, Faculdade de Ci\^encias, UNESP, C.P. 473, 17033-360, Bauru, SP, Brazil}

\author{G. G. Lesseux}
\affiliation{1. Physikalisches Institut, Universit\"at Stuttgart, D-70569, Germany}

\author{H. B. Pizzi}
\affiliation{Instituto de F\'isica \lq\lq Gleb Wataghin\rq\rq, UNICAMP, 13083-859, Campinas, SP, Brazil}

\author{G. S. Freitas}
\affiliation{Instituto de F\'isica \lq\lq Gleb Wataghin\rq\rq, UNICAMP, 13083-859, Campinas, SP, Brazil}
\affiliation{Los Alamos National Laboratory, Los Alamos, New Mexico 87545, USA}

\author{R. R. Urbano}
\affiliation{Instituto de F\'isica \lq\lq Gleb Wataghin\rq\rq, UNICAMP, 13083-859, Campinas, SP, Brazil}

\author{P. A. Venegas}
\affiliation{Departamento de F\'isica, Faculdade de Ci\^encias, UNESP, C.P. 473, 17033-360, Bauru, SP, Brazil}

\author{P. G. Pagliuso}
\affiliation{Instituto de F\'isica \lq\lq Gleb Wataghin\rq\rq, UNICAMP, 13083-859, Campinas, SP, Brazil}
\affiliation{Los Alamos National Laboratory, Los Alamos, New Mexico 87545, USA}


\date{\today}

\begin{abstract}
The gap opening mechanism of a topological insulator, the quantum anomalous Hall effect and the axion physics are still pressing open questions and a microscopic viewpoint to further understand the role of magnetism in topology is highly desirable. In this work we have performed a microscopic investigation, by means of electron spin resonance (ESR) along with complementary bulk measurements, on the chalcogenide (Bi$_{1-x}$Gd$_{x}$)$_{2}$Se$_{3}$ ($x$ = 0, 0.001, 0.002 and 0.006). Our analysis of the Gd$^{3+}$ spin dynamics reveal no significant change of the Fermi surface as a function of Gd$^{3+}$ concentration, which indicates that the 4$f$ magnetism is different from the non-local effects induced by transition metals ($d$ electrons) substitutions. Additionally, we observe an unusual evolution of the Gd$^{3+}$ ESR spectra as a function of the applied magnetic field, which we discuss considering the magnetic interaction between Gd$^{3+}$ 4$f$ electrons and impurity centers such as Se vacancies. This interaction would give rise to a local weak antilocalization effect surrounding the Gd$^{3+}$ ions. Such mechanism is observable due to particular details of the Gd$^{3+}$ 4$f$ electrons magnetism in this system compared to $d$ electrons. Our work points out that rare earth substitutions in this model topological insulator is a promising path to explore the axion insulating systems.
\end{abstract}

\pacs{76.30.-v, 71.20.Lp}
\maketitle

\section{\label{sec:intro}I. Introduction}

The application of topology in condensed matter physics has been responsible to unveil new quantum phases of matter, the topological insulators (TIs) in two \cite{bernevig2006quantum,konig2007quantum,culcer2020transport} and three dimensions \cite{hasan2010colloquium,culcer2020transport} being the most prominent and heavily explored systems. The insulating bulk with protected spin-polarized gapless surface states is a highly attractive characteristic of TIs \cite{hasan2010colloquium,tokura2017emergent}. Another interesting property rises from the interplay between topology and magnetism, where the time reversal symmetry breaking could result in axion physics and the quantum anomalous hall effect (QAHE) \cite{zhang2012interplay,chang2013experimental,he2014quantum,chang2015high,grauer2015coincidence,xiao2018realization,yue2019symmetry,zhang2019topological,deng2020quantum,liu2020robust,nenno2020axion,fijalkowski2021any}.

The (Bi,Sb)$_{2}$(Te,Se)$_{3}$ chalcogenides have been established as the model systems of three dimensional TIs due to the single Dirac cone near the Fermi level \cite{zhang2009topological,analytis2010bulk,pan2011electronic}. Naturally, such systems have been able to host the interplay between magnetism and topology \cite{chang2013experimental,he2014quantum,chang2015high}. The first realization of the QAHE was achieved in (V,Cr):(Bi,Sb)$_{2}$Te$_{3}$ thin films, however it is possible to obtain a fully quantized Hall conductivity only at milikelvin temperatures \cite{chang2013experimental,he2014quantum,chang2015high,grauer2015coincidence}. This limitation may be due to the presence of thermally activated bulk carriers, which cause a breakdown of the QAHE \cite{fijalkowski2021quantum}. One remaining question relies on the role of the underlying magnetic coupling mechanism. More specifically, it is imperative to understand how the QAHE can be affected by more complex magnetic interactions which goes beyond the so-called van Vleck mechanism. This question has recently been explored by electron spin resonance (ESR), x-ray absorption and resonant photoelectron spectroscopy \cite{mahani2014interplay,zimmermann2016spin,peixoto2016impurity,zhang2018electronic,ye2019negative,bouaziz2019spin,tcakaev2020comparing,peixoto2020non}. Indeed the $p-d$ hybridization, along with the $d$ states occupation and the consequent Sb and Te polarization, near the Fermi level seems to play an important role to the magnetism of substituted chalcogenides, which goes beyond the van Vleck mechanism \cite{zhang2018electronic,islam2018systematics,ye2019negative,peixoto2020non}. Exploring heterostructures is also a promising, yet challenging, path to achieve higher temperatures in the QAHE \cite{fijalkowski2021any,mogi2017magnetic,mogi2017tailoring,xiao2018realization,mathimalar2020signature,pereira2020topological,pereira2020interfacing}.

One little explored but promising route to understand the role of the magnetism and its influence in the bulk is to investigate the magnetism of 4$f$ electrons. For instance, Sm-substituted chalcogenides could display higher order topological insulators phases with chiral hinge states \cite{chen2015high,yue2019symmetry}. Europium substituted chalcogenides shows antiferromagnetic correlations, however Eu has a 2+ valence, while most of the rare-earths shows a 3+ state \cite{tcakaev2020incipient}. Angle resolved photoemission spectroscopy studies show the robustness of the surface states to a Gd$^{3+}$ concentration of $\sim$ 0.1 in Bi$_{2}$Te$_{3}$ and TlBiSe$_{2}$ \cite{li2013magnetic,filnov2020probe}. Therefore, further exploring 4$f$-substituted systems can be highly interesting to understand the gap opening mechanism and the axion insulating phase in these model systems \cite{lee2019angle,lee2015antiferromagnetic,kim2018gd,filnov2020probe,filnov2019magnetic}.

In this context, Gd$^{3+}$, which also induces antiferromagnetic correlations \cite{kholdi1994magnetic,song2012large,kim2018gd}, is an ideal testbed due to the stable valence and the weak crystalline electrical field (CEF) effects. Although Gd$^{3+}$-substituted chalcogenides have been explored in macroscopic \cite{kholdi1994magnetic,song2012large} and previous ESR studies \cite{kholdi1994magnetic,isber1995hyperfine,gratens1997epr,garitezi2015electron}, a detailed investigation of the 4$f$ local magnetic effects and the spin dynamics induced by these substitutions, as well as a comparison to more traditional $d$ systems is missing \cite{bouaziz2019spin,zimmermann2016spin}.

In this work we locally explore Gd$^{3+}$-substituted Bi$_{2}$Se$_{3}$ using ESR at different frequencies ($\nu$ = 9.5 and 34 GHz). We show that the introduction of Gd$^{3+}$ ions does not alter the carriers near the Fermi surface, which are mainly $p$ states. This is a different mechanism from the substitution using ions where the magnetism comes from $d$ states. Such difference is also manifested into the Gd$^{3+}$ spin dynamics and in the macroscopic properties of the system. Additionally, the Gd$^{3+}$ ESR data show an unusual evolution as a function of the applied magnetic field. While for lower field we obtain a Gd$^{3+}$ response with resolved fine structure, which is more consistent with Gd$^{3+}$ ions in an insulating environment, at higher fields the system tends to behave as a conventional metal, revealing a single additional line with collapsed fine structure. We discuss the evolution of the Gd$^{3+}$ local environment under the light of a possible local weak antilocalization (WAL) effect \cite{hikami1980spin,kim2011thickness,lu2011weak,lu2011competition}, which is a product of the interplay between strong spin-orbit coupling and the interaction between 4$f$ local moments and impurity centers such as Se vacancies. Our results shed light on the microscopic mechanism involving the introduction of 4$f$ electrons into the chalcogenides. 

\section{\label{sec:experiment}II. Methods}

The chalcogenides have a rhombohedral crystal structure. Single crystalline samples of (Bi$_{1-x}$Gd$_{x}$)$_{2}$Se$_{3}$ ($x$ = 0, 0.001, 0.002 and 0.006) were grown by the stoichiometric melting technique. High purity Bi, Gd and Se elements were put inside of an alumina crucible with the ratio of (2 - $2x$):$2x$:3. The crucible was vacuum sealed in a quartz tube, heated to 800$^{\circ}$C for 72 hours and cooled down to room temperature at 2$^{\circ}$C per hour. We cut the crystals in a rectangular-like shape. Typical sample sizes were 0.5 x 2 x 0.3 mm$^{3}$. The structure and phase purity were checked by x-ray powder diffraction using a commercial diffractometer (Cu-K$\alpha$), where we confirmed the single phase nature of our samples. We have performed elemental analysis using energy and wavelenght dispersive spectroscopy, which revealed a small, but measurable, amount of Se vacancies ($\sim$ 0.1). Magnetic susceptibility measurements were performed in a commercial SQUID magnetometer. Specific-heat measurements were done in a commercial small-mass calorimeter system. Electrical resistivity data were acquired using a four-probe configuration with a $dc$ resistance bridge. ESR measurements were performed on single crystals in X- and Q-band ($\nu$ = 9.5 and 34 GHz respectively) spectrometer equipped with a goniometer and a He-flow cryostat in the temperature range of 4 K $\leq$ $T$ $\leq$ 300 K at low power $P$ $\leq$ 2 mW. The ESR spectra were analyzed using the software SPEKTROLYST.

\section{\label{sec:results}III. Results}

\begin{figure}[!ht]
\includegraphics[width=0.99\columnwidth]{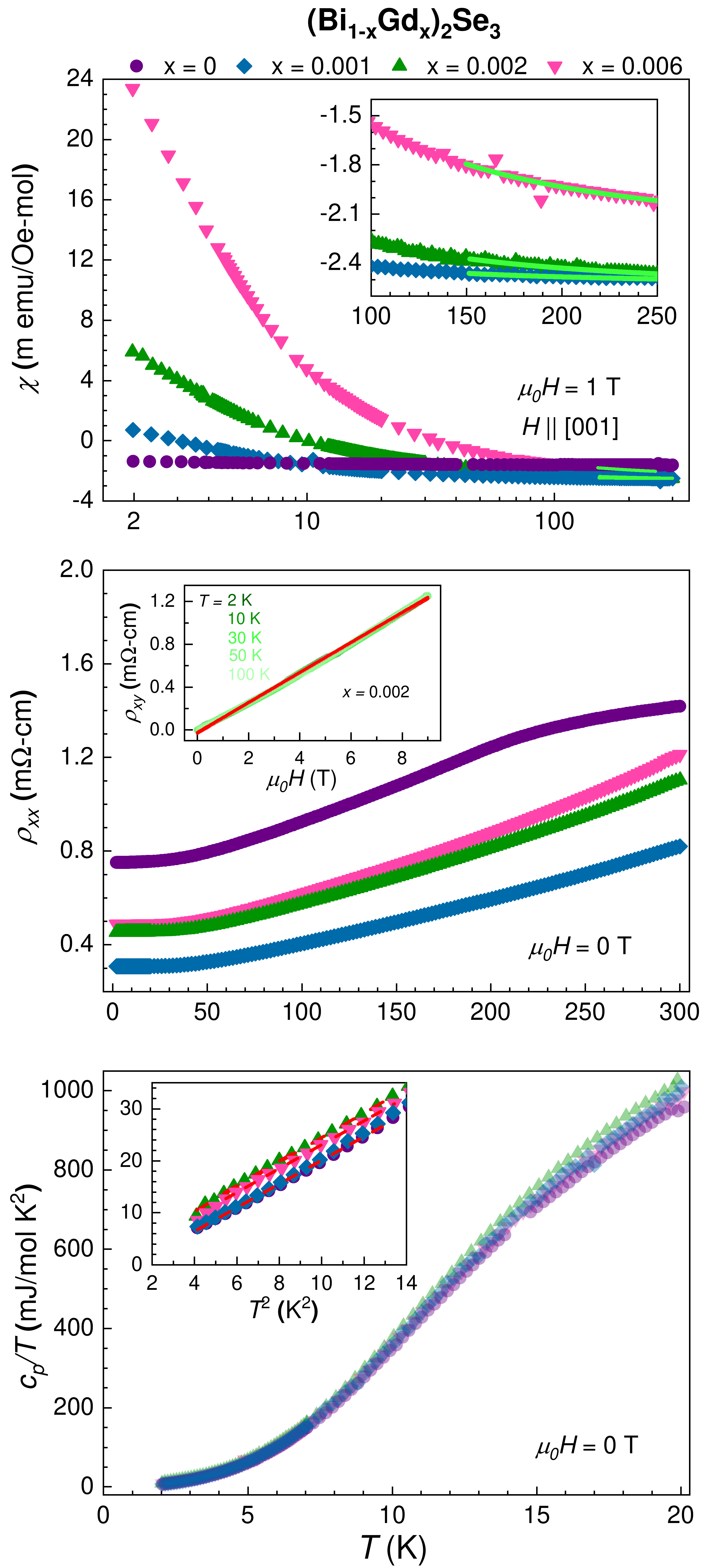}
\caption{(Top panel) Magnetic susceptibility for $H$ applied parallel to the [001] direction, (middle panel) longitudinal resistivity and (lower panel) specific heat as a function of temperature for (Bi$_{1-x}$Gd$_{x}$)$_{2}$Se$_{3}$. The insets show the (top panel) susceptibility at high temperatures, (middle panel) Hall resistivity as a function of applied magnetic field for different temperatures for $x$ = 0.002, and (bottom panel) the specific heat divided by temperature as a function of $T^{2}$ at low temperatures.}
\label{Fig1}
\end{figure}

Figure \ref{Fig1} summarizes the macroscopic properties of (Bi$_{1-x}$Gd$_{x}$)$_{2}$Se$_{3}$. The top panel shows the magnetic susceptibility as a function of temperature. From a Curie-Weiss fitting $\chi = \chi_{0} + C/(T-\theta$), where $\chi_{0}$ is the $T$-independent term, $C$ is a constant and $\theta$ the Curie-Weiss temperature, at high temperatures (150 K $\leq$ $T$ $\leq$ 300 K) we can estimate the concentration of Gd$^{3+}$ in (Bi$_{1-x}$Gd$_{x}$)$_{2}$Se$_{3}$. Assuming 7.94 $\mu_{B}$/Gd, we obtained $x$ $\approx$ 0.001, 0.002 and 0.006 with a $\theta$ = -1(5) K for all samples. Taking into account the nuclear diamagnetism, we can also estimate the Pauli magnetic susceptibility $\chi_{p}$ = 20(10) $\mu$emu/mol Oe for all samples.

The middle panel shows the longitudinal resistivity $\rho_{xx}$ as a function of temperature for (Bi$_{1-x}$Gd$_{x}$)$_{2}$Se$_{3}$. We obtain a metallic-like behavior, which is expected due to the presence of Se vacancies in single crystals \cite{kim2011thickness,huang2012nonstoichiometric,devidas2015role,adroguer2015conductivity}. Nonetheless, we only observe small and not systematic differences between samples, which can be attributed to a small variation of Se vacancies from crystal to crystal. As a result, the residual resistivity $\rho_{0}$, which can be associated with disorder in the system, can differ between samples from different batches. The lack of a systematic change indicates that Gd$^{3+}$ substitution at the Bi site does not introduce carriers into the system, and presumably its 4$f$ electrons have no relevant role in the Fermi surface. This is in contrast to substitutions using transition metal ions where the magnetism originates from $d$ orbitals \cite{teng2019mn,zimmermann2016spin,peixoto2020non,janivcek2008transport}. The absence of Gd$^{3+}$-introduced carriers into the system as a function of Gd$^{3+}$ concentration is also supported by the Hall resistivity ($\rho_{xy}$), which is Gd$^{3+}$ concentration and temperature independents, as shown in the inset of the middle panel. The Hall response to the applied magnetic field is linear and positive, consistent with the transport properties being dominated by a single band of holes ($p$-type). From a linear fit ($\rho_{xy}$ = 1/ne, where $e$ is the electron charge) we can extract a carrier density of $n_{h}$ = 5(3).10$^{18}$ h/cm$^{3}$.

The bottom panel shows the specific heat divided by the temperature $c_{p}$/$T$ as a function of temperature for (Bi$_{1-x}$Gd$_{x}$)$_{2}$Se$_{3}$. There is a slight difference between different Gd$^{3+}$ concentrations, which reinforces the lack of change in the Fermi level due to the Gd$^{3+}$ substitution. Performing linear fits for $c_{p}$/$T$ as a function of $T^{2}$, top inset of the bottom panel, it is possible to extract the Sommerfeld coefficient $\gamma$ = 0.8(3) mJ/mol K$^{2}$ for all samples. For a free conduction electron gas model $\gamma = (2/3) \pi k_{B}^{2} \eta(E_{F})$, where $k_{B}^{2}$ is the Boltzmann constant and $\eta(E_{F})$ the density of states (DOS) at the Fermi level per spin. We extract $\eta(E_{F})$ = 0.16(6) states/eV mol spin for all samples. We can further analyze the role of electron-electron ($ee$) interactions in the bulk by comparing the estimated Pauli susceptibility ($\chi_{p}^{theor}$ = 2$\mu_{B}^{2}\eta(E_{F})$, where $\mu_{B}$ is the Bohr magneton) with the experimental value. We obtain $\chi_{p}^{theor}$ = 15 $\mu$emu/mol Oe. These results indicate that the DOS at the Fermi level is not affected as a function of Gd$^{3+}$ concentration and, if any, the role of $ee$ interactions is negligible in the bulk. It is noteworthy that recent results point out that $ee$ correlations do play a role in the surface states of chalcogenides \cite{chen2011tunable,wang2011evidence,pal2012effect}.

\begin{figure}[!ht]
\includegraphics[width=0.99\columnwidth]{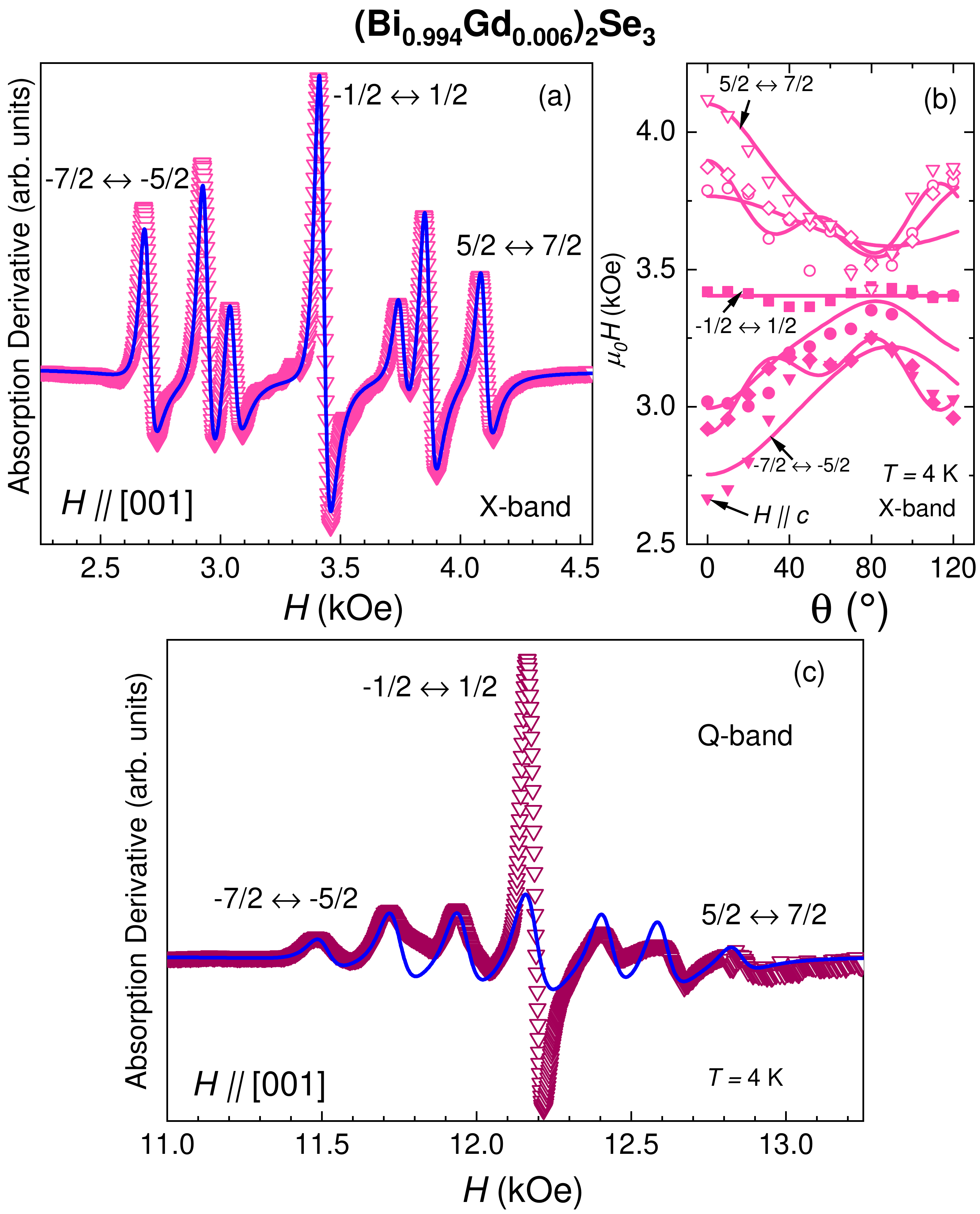}
\caption{(a) X-band ($\nu$ = 9.5 GHz) Gd$^{3+}$ ESR spectra for (Bi$_{0.994}$Gd$_{0.006}$)$_{2}$Se$_{3}$ and (b) its resonance field angle dependence at $T$ = 4 K with the applied magnetic field $H$ parallel to the [001] direction. In (b) the field is rotated towards the $ab$ plane. (c) Q-band ($\nu$ = 34 Ghz) Gd$^{3+}$ ESR spectra. The blue solid lines are simulations considering the relative intensities for each of the seven Gd$^{3+}$ fine structure lines. The magenta solid lines are fits considering the hexagonal CEF Hamiltonian described into the text.}
\label{Fig2}
\end{figure}

As mentioned earlier, Gd$^{3+}$ is a $S$-ion ($L$ = 0, $S$ = 7/2). As such, the CEF effects appears only due to a small mixing of excited states in the ground state, resulting in a intermediate coupling \cite{abragam2012electron,barnes1981theory}. Due to this small correction, the CEF splitting is of the order of the Zeeman energy and we are able to observe the Gd$^{3+}$ fine structure, with a selection rule of $\Delta m_{s}$ = $\pm$ 1 \cite{abragam2012electron,barnes1981theory}. Another consequence of the $S$-ion nature is that the $g$-value of an isolated Gd$^{3+}$ ion is independent of the symmetry of the matrix, which makes the analysis of the Gd$^{3+}$ spin dynamics more straightforward \cite{abragam2012electron,barnes1981theory}. With that in mind, in order to have a microscopic insight about the local effects of Gd$^{3+}$ ions, we have performed ESR in two different frequencies. Figure \ref{Fig2} shows the Gd$^{3+}$ ESR spectra at $T$ = 4 K for $x$ = 0.006 with the applied field $H$ parallel to the [001] direction and its angle dependence as an example. Fig. \ref{Fig2}(a) shows the Gd$^{3+}$ X-band ESR spectrum, where we can observe the resolved Gd$^{3+}$ fine structure displaying seven transition lines \cite{abragam2012electron,barnes1981theory,poole1971relaxation} - with their expected relative intensities. Such observation excludes the possibility of Gd$^{3+}$ ions having distinct sites and/or being interstitial in Bi$_{2}$Se$_{3}$. It also corroborates that we only have one phase in our crystals. Each individual resonance line can be described as a Dysonian line shape, which is typical when the skin depth $\delta$ is smaller than the sample size $d$ \cite{feher1955electron,dyson1955electron}. Indeed, from the resistivity measurements shown in Fig. \ref{Fig1}(b), we can estimated for X-band $\delta$ $\sim$ 11 $\mu$m ($\ll$ $d$ $\sim$ 300 $\mu$m for our samples) at $T$ = 4 K, consistent with the Dysonian line shape. Each individual line, represented by the power absorption derivative d$P$/d$H$ as a function of $H$, can be described as an admixture of absorption and dispersion derivatives 

\begin{equation}
\frac{dP}{dH} \propto (1-\lambda)\frac{\mathrm{d} }{\mathrm{d} x}\left ( \frac{1}{1+x^2} \right 
) + \lambda \frac{\mathrm{d} }{\mathrm{d} x} \left ( \frac{x}{1+x^2} \right ),
\label{Eq1}
\end{equation}
where $\lambda$ is the asymmetric parameter of the line shape and $x$ = $2(H - H_{r})/\Delta H$ \cite{feher1955electron}, wherein $H_{r}$ is the Gd$^{3+}$ resonance field and $\Delta H$ the linewidth.

As already mentioned, the position of the Gd$^{3+}$ fine structure is dependent on the local symmetry of Gd$^{3+}$ ions. In the case of the Bi$_{2}$Se$_{3}$, the local symmetry is hexagonal, where the spin Hamiltonian is given by

\begin{equation}
\begin{split}
\mathcal{H}= & \frac{1}{3} b_{2}^{0}O_{2}^{0}+ \frac{1}{60} \left ( b_{4}^{0}O_{4}^{0}+b_{4}^{3}O_{4}^{3} \right ) + \\ 
 & \frac{1}{1260} \left ( b_{6}^{0}O_{6}^{0}+b_{6}^{3}O_{6}^{3}+b_{6}^{6}O_{6}^{6} \right ) +g\mu_{B}\mathbf{H}\mathbf{S},
\label{EqHamiltonian}
\end{split}
\end{equation}
where $b_{n}^{m}$ are the $n$ order CEF parameters and $O_{m}^{n}$ the Stevens operator. The Gd$^{3+}$ fine structure resonance fields are going to be highly angular dependent \cite{abragam2012electron,barnes1981theory}. Therefore, we fitted the angular dependence of the Gd$^{3+}$ fine structure resonance fields with Eq. \ref{EqHamiltonian}, as shown by the solid magenta lines in Fig. \ref{Fig2}(b). The best parameters obtained were $b_{2}^{0}$ = 37.5(5) Oe, $b_{4}^{3}$ = 0.03(2) Oe and $b_{6}^{6}$ = - 0.3(2) Oe. The values of $b_{4}^{0}$, $b_{6}^{0}$ and $b_{6}^{3}$ were negligible. The dominant term, which is the axial $b_{2}^{0}$, is consistent with previous CEF studies of Gd$^{3+}$ in Bi$_{2}$Se$_{3}$ \cite{gratens1997epr}. Albeit the contribution from smaller terms may change from this previous study to our result, it is important to observe that the absolute signs of $b_{2}$ $>$ 0, $b_{4}$ $>$ 0 and $b_{6}$ $<$ 0 are in agreement between both studies \cite{gratens1997epr}. The change between which terms are negligible from one to another may be due to those terms being much smaller than the dominant one, and we can obtain a local minimum different for each rotation. Nonetheless, our obtained CEF parameters are in accordance with \cite{gratens1997epr}. The blue solid line is a simulation considering the value of $b_{2}^{0}$ of this hexagonal CEF Hamiltonian.

At this point it is interesting to compare our X-band Gd$^{3+}$ spectrum in Bi$_{2}$Se$_{3}$ with previous reports of Gd$^{3+}$ and Mn$^{2+}$ in Bi$_{2}$Te$_{3}$ \cite{zimmermann2016spin,isber1995hyperfine,kholdi1994magnetic}. While both (X- and Q-bands) Gd$^{3+}$ spectra show resolved fine structure, the same does not happen for Mn$^{2+}$, in which only one resonance is observed even for the smallest Mn$^{2+}$ concentration ($\sim$ 0.005) \cite{zimmermann2016spin}. Mn$^{2+}$ is a $S$ = 5/2 ion, therefore there is an exchange narrowing of the fine structure going from $d$ to $f$ states, which reflects different substitution effects. The exchange narrowing will result from the large exchange interaction from $d$ states with the $p$ polarized states in these systems \cite{teng2019mn,zimmermann2016spin,peixoto2020non,anderson1953exchange,urban1975narrowing}.

Fig. \ref{Fig2}(c) shows the Gd$^{3+}$ Q-band ESR spectrum at $T$ = 4 K. The blue solid line is a simulation of the same CEF scheme of the X-band data, normalized by the intensity of the Gd$^{3+}$ fine structure. There is a clear difference when compared to the X-band data, with a stronger - 1/2 $\leftrightarrow$ 1/2 transition, which we will further call Gd$^{3+}$ central line, at the top of the seven fine structure lines. Such coexistence can be linked to two different environments for Gd$^{3+}$ ions distributed along the sample. Although the distribution of the Gd$^{3+}$ ions appears to be structurally homogeneous, due to the X-band data, the electronic environment around the localized moments seems to show an interesting evolution as a function of the applied magnetic field. Such effect may be connected with the Se vacancies distribution, as we discuss below analyzing the Gd$^{3+}$ spin dynamics in more details.
 
\begin{figure}[!ht]
\includegraphics[width=0.99\columnwidth]{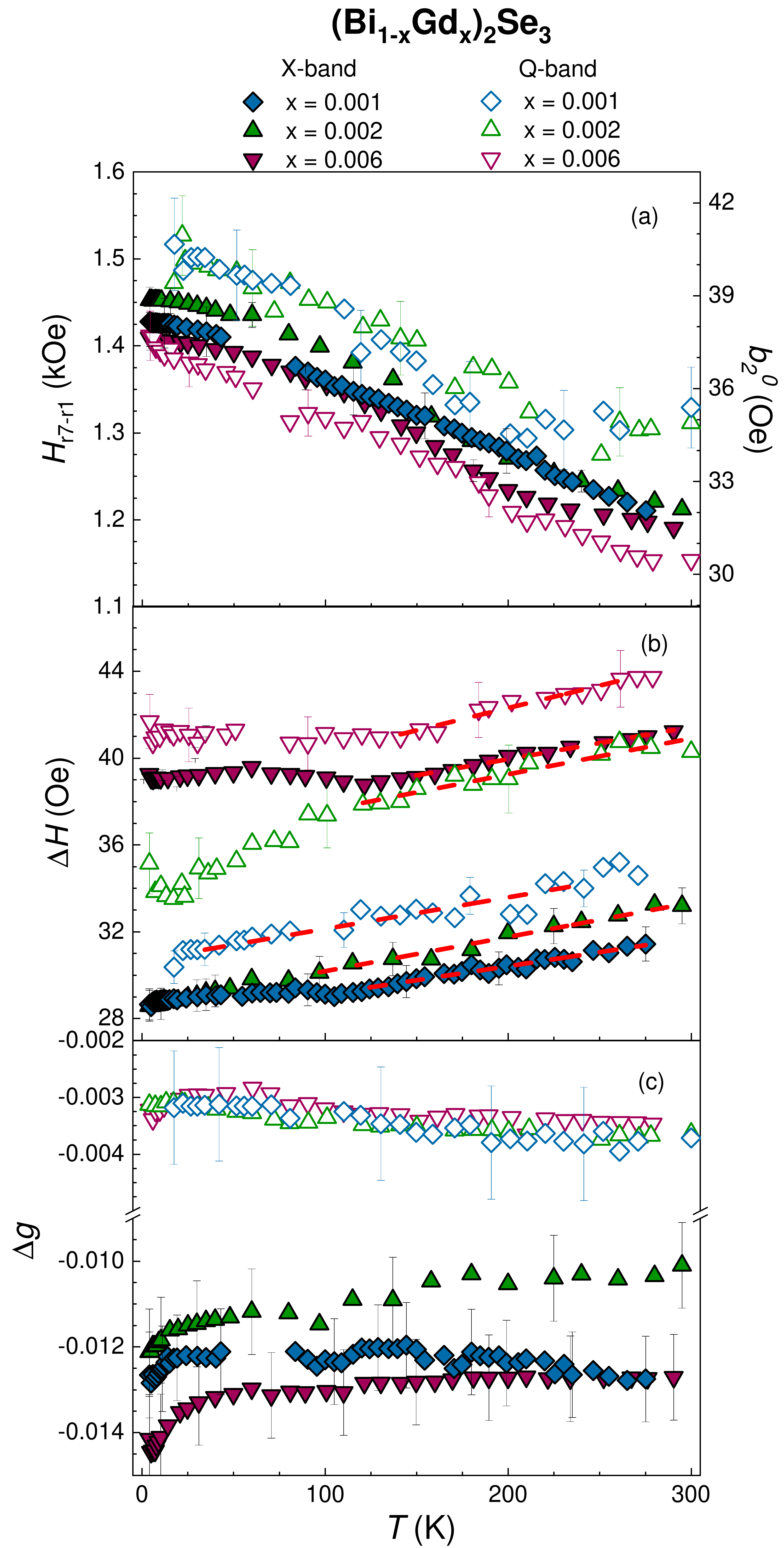}
\caption{(a) Gd$^{3+}$ spectrum splitting and the axial term $b_{2}^{0}$, (b) $\Delta H$ and (c) $g$-shift as a function of temperature for (Bi$_{0.994}$Gd$_{0.006}$)$_{2}$Se$_{3}$ for the applied magnetic field parallel to the [001] direction. $\Delta H$ and $g$-shift were extracted from the ESR central line. The red dashed lines in (b) represent the linear fits used to extract the Korringa rates. We used $g_{theor}$ = 1.993(1) in order to calculate the $g$-shift for all samples.}
\label{Fig3}
\end{figure}

The Gd$^{3+}$ ESR spectrum temperature evolution and the spin dynamics can bring up valuable information about the introduction of 4$f$ states in Bi$_{2}$Se$_{3}$ and their field dependence. Figure \ref{Fig3} summarizes this temperature evolution. The ESR fine structure split, defined by the difference between the resonance field $H_{r}$ of the $\mp$ 7/2 $\leftrightarrow$ $\mp$ 5/2 transitions ($H_{r7}$ - $H_{r1}$), is shown in Fig. \ref{Fig3}(a). Since the axial term of the CEF parameters is the most dominant one, $b_{2}^{0}$ is going to have to be directly proportional to the CEF splitting. We also show the value of $b_{2}^{0}$ for each CEF splitting in Fig. \ref{Fig3}(a). The Gd$^{3+}$ splitting is nearly constant ($\Delta_{H_{r7} - H_{r1}}^{4 - 50 K}$/$T$ $\leq$ 0.4 Oe/K) until $\sim$ 50 K for both bands. At higher temperatures we observe a more significant ($\Delta_{H_{r7} - H_{r1}}^{50 - 300 K}$/$T$ $\sim$ 1 Oe/K) and systematic reduction of the fine structure splitting. This narrowing can be related with the interaction between carriers or, more likely, due to changes in the CEF effects related to the thermal expansion of the compound \cite{gratens1997epr}. Indeed, previous neutron diffraction and pair density function analysis have shown a local anharmonic thermal expansion in Bi$_{2}$Se$_{3}$ \cite{park2013local}. In particular, the two Se sites have a distinct thermal expansion in reference to the Bi site \cite{park2013local}. Therefore, this anharmonic expansion should affect the CEF effects in the Bi site and, consequently, should influence the Gd$^{3+}$ CEF splitting. Our results are in agreement with previous results \cite{gratens1997epr,garitezi2015electron} and shows the similarity of the fine structure between X- and Q-bands spectra.

From the analysis using an admixture of absorption and dispersion derivatives using Eq. \ref{Eq1} one can extract the Gd$^{3+}$ $\Delta H$ and $H_{rc}$ of the central line. As a result, we can also obtain the experimental Gd$^{3+}$ $g$-value $g_{exp}$ = $h\nu/\mu_{B} H_{rc}$, where $h$ is the Planck constant and $\mu_{B}$ the Bohr magneton. Figures \ref{Fig3}(b) and \ref{Fig3}(c) show the Gd$^{3+}$ linewidth $\Delta H$ of the central line and $g$-shift $\Delta g$ = $g_{exp}$ - $g_{theor}$ as a function of temperature. Here $g_{theor}$ = 1.993(1), which is the Gd$^{3+}$ $g$-value in an insulating matrix. The Gd$^{3+}$ central line is a reliable resonance to analyze the trend of the Gd$^{3+}$ spin dynamics due to a smaller influence of CEF effects in the $g$-value \cite{barnes1981theory}.

Regarding the $\Delta H$ of the Gd$^{3+}$ central line, all the obtained data show a linear increase at high temperatures ($T$ $\geq$ 80 K). We focused our analysis on the high-$T$ region to avoid the contribution of possible Gd$^{3+}$-Gd$^{3+}$ interactions at low-$T$, especially for samples with higher concentrations and higher fields. Such linear increase can be attributed to a relaxation process through the exchange interaction between the Gd$^{3+}$ 4$f$ local moments and the carriers, which eventually results in the spin-flip scattering of the latter. This spin-spin relaxation mechanism is known as Korringa relaxation \cite{abragam2012electron,barnes1981theory,poole1971relaxation}. From a linear fit $\Delta H$ = $\Delta H_{0}$ + $bT$ we can extract the Gd$^{3+}$ residual linewidth $\Delta H_{0}$ and the Korringa rate $b$. The results are summarized in Table \ref{TableI}. While $b$ is related to the spin-spin relaxation, $\Delta H_{0}$ can be associated to, e.g., disorder and sample homogeneity \cite{abragam2012electron,barnes1981theory,poole1971relaxation}.

\begin{table*}[!ht]
  \centering
	\setlength{\extrarowheight}{2.5pt}
	\caption{ESR parameters extracted from the Gd$^{3+}$ spin dynamics analysis for (Bi$_{1-x}$Gd$_{x}$)$_{2}$Se$_{3}$.}
  \begin{tabular}{C{1.4cm}C{1.4cm}C{1.4cm}C{2.5cm}C{1.4cm}C{1.4cm}C{1.4cm}C{2.5cm}C{1.4cm}}
    \hline
		& \multicolumn{4}{c}{X-band} & \multicolumn{4}{c}{Q-band} \\
    & $\Delta H_{0}$ & $b$ & $\langle J_{fp}^{2}(\textbf{q}) \rangle^{1/2}$ & $J_{fp}(0)$ & $\Delta H_{0}$ & $b$ & $\langle J_{fp}^{2}(\textbf{q}) \rangle^{1/2}$ & $J_{fp}(0)$ \\
		& Oe & Oe/K & meV & meV & Oe & Oe/K & meV & meV \\
    \hline
		$x$ = 0.006 & 37(2) & 0.015(5) & $\sim$ 5 & $\sim$ 80 & 38(2) & 0.020(5) & $\sim$ 5 & $\sim$ 20 \\
    $x$ = 0.002 & 29(2) & 0.016(5) & $\sim$ 5 & $\sim$ 80 & 36(2) & 0.017(5) & $\sim$ 5 & $\sim$ 20 \\
    $x$ = 0.001 & 28(2) & 0.013(5) & $\sim$ 5 & $\sim$ 80 & 31(2) & 0.015(5) & $\sim$ 5 & $\sim$ 20 \\
    \hline
  \end{tabular}
	\label{TableI}
\end{table*}

Turning our attention to the Gd$^{3+}$ X-band results, $b$ is, within experimental uncertainty, the same for all concentrations, which indicates the absence of exchange bottleneck effects \cite{abragam2012electron,barnes1981theory,poole1971relaxation}. $\Delta H_{0}$ increases systemically as a function of Gd$^{3+}$ concentration, which is expected due to the increase of disorder in the system. At low temperatures we see a deviation of the Korringa rate, which might be associated with the interaction between local moments and impurity centers. As such, as mentioned earlier, a proper analysis is to focus on the high temperature Gd$^{3+}$ spin dynamics data. Looking now to the field dependency, we still obtain the same Korringa rate for all concentrations [Fig. \ref{Fig3}(b)]; however $\Delta H_{0}$ increases systematically when comparing X- and Q-bands results for the same concentration. This is an indication of inhomogeneous broadening, which raises most likely from slightly different CEF states around the Gd$^{3+}$ sites \cite{abragam2012electron,barnes1981theory,poole1971relaxation}. Bi$_{2}$Se$_{3}$ is well known to host Se vacancies intrinsically, which can show a small inhomogeneity across the sample \cite{kim2011thickness,xu2014intrinsic,tayal2017role}. Moreover, previous nuclear magnetic resonance measurements have shown that, in polycrystalline Bi$_{2}$Se$_{3}$, defect regions segregate into domains \cite{taylor2012spin}.

Before analyzing the spin relaxation of our Gd$^{3+}$ probe in more details, it is instructive to first look to the $g$-shift as a function of temperature, applied magnetic field and Gd$^{3+}$ concentration - Figure $\ref{Fig3}$(c). On one hand, $s$ and/or $d$ carriers have a ferromagnetic (atomic) interaction with 4$f$ local moments, which produces a positive $g$-shift. On the other hand, $p$ and/or $f$ carriers magnetic interaction with 4$f$ local moments occurs through the so-called virtual bound states \cite{davidov1973crystalline}, which results in an antiferromagnetic (covalent) interaction between them. In this second case, the result is a negative $g$-shift. Therefore, the sign of the $g$-shift is crucial to have information about the nature of the wave function of the carriers of the system.

As such, an important result reported here is the negative $g$-shift, confirming $p$ states as the main carriers near the Fermi level \cite{zhang2009topological}. This result is consistent with previous angle resolved spectroscopy studies \cite{cao2013mapping,vidal2013photon}. Looking more specifically to the X-band data, at low temperatures we can see a systematic decrease of the $g$-shift for all concentrations. This is due to an antiferromagnetic interaction, which is in accordance with the observed AFM ground state in Eu$^{2+}$ and Gd$^{3+}$-substituted chalcogenides \cite{kim2015antiferromagnetic,tcakaev2020incipient}. However, such decrease is observed even for samples with a concentration as low as $x$ = 0.001, further suggesting that such contribution comes from a possible interaction between 4$f$ local moments and impurity centers, such as Se vacancies, which also have a signature in the Gd$^{3+}$ $\Delta H$ $T$-dependence - Fig. \ref{Fig3} (b). Similar contributions recently observed in organic salts were also interpreted to have origin in impurity centers \cite{shimizu2006emergence,riedl2019critical,kawamura2019nature,pustogow2020impurity,miksch2021gapped}. At high temperatures, the Gd$^{3+}$ $\Delta g$ values are $T$-independent within experimental uncertainty, which shows that dynamic effects are negligible into the systems \cite{rettori1974dynamic}. While the Gd$^{3+}$ $g$-values for $x$ = 0.001 and 0.006 are virtually identical, as expected, for $x$ = 0.002 we observe a subtle, but systematic, increase of the $g$-value. Such small difference might also be related with samples with $x$ = 0.002 showing a higher quantity of impurity centers, such as Se vacancies. This interpretation is corroborated by a few points: The first one is that the increase of $\Delta H_{0}$ for $x$ = 0.002 going from X- to Q-band is relatively larger when compared to other Gd$^{3+}$ concentrations. Secondly, the deviation at low temperatures of the linear increase in the Gd$^{3+}$ $\Delta H$ are more pronounced for $x$ = 0.002 when compared to other concentrations. Finally, the $g$-shift for all three concentrations are the same for Q-band measurements, which indicates that there is no change in $\eta(E_{F})$ as a function of Gd$^{3+}$ concentration, consistent with specific heat measurements.

Interestingly, there is also a decrease of the negative Gd$^{3+}$ $g$-shift as a function of the applied magnetic field for all concentrations of (Bi$_{1-x}$Gd$_{x}$)$_{2}$Se$_{3}$. In the absence of dynamic, bottleneck and multiple band effects, $b$ and $\Delta g$ can be described as

\begin{equation}
b = \frac{\pi k_{B}}{g\mu_{B}} \langle J_{fp}^{2}(\textbf{q}) \rangle \eta^{2} (E_{F})\frac{K(\alpha)}{(1 - \alpha)^{2}},
\label{Eq2}
\end{equation}

\begin{equation}
\Delta g = J_{fp}(0)\, \frac{\eta (E_{F})}{(1 - \alpha)},
\label{Eq3}
\end{equation}
where $k_{B}$ is the Boltzmann constant, $J_{fp}(0)$ is the effective exchange interaction between the Gd$^{3+}$ local moments and the carriers for the momentum transfer $q$ = 0, and $\langle J_{fp}^{2}(\textbf{q}) \rangle^{1/2}$ is the average of the exchange interaction with momentum transfer $q$ at the residual Fermi surface \cite{abragam2012electron,barnes1981theory,poole1971relaxation}. Any possible relevant $ee$ correlations are taken into account in the Stoner enhancement factor $(1-\alpha)^{-1}$ \cite{moriya1963effect,narath1967nuclear} and in the Korringa exchange factor $K(\alpha)$ \cite{narath1968effects,shaw1971enhancement}. As already shown, the $ee$ correlations in the bulk does not seem to play an important role, therefore we assume $\alpha$ = 0 and $K(\alpha)$ = 1. The estimated exchange interactions for each Gd$^{3+}$ concentration and applied magnetic field are summarized in Table \ref{TableI}. Although we should underestimate $J_{fp}(0)$ and $<J_{fp}^{2}(\textbf{q})>^{1/2}$ due to remaining CEF effects and a local reduction of the DOS, the trend observed in our analysis qualitatively trustworthy.

\section{\label{sec:discussion}IV. Discussion}

As clearly shown in Table \ref{TableI}, in X-band measurements we do have a clear $\textbf{q}$-dependence in the exchange interaction for (Bi$_{1-x}$Gd$_{x}$)$_{2}$Se$_{3}$, which appears to reduce upon increasing the applied magnetic field (Q-band). A notion of this effect can be obtained looking to the real space. Our results indicate that, at lower fields, there is a stronger exchange interaction surrounding the Gd$^{3+}$ ions. In other words, it appears that occurs a localization of the carriers surrounding the Gd$^{3+}$ ions. The increase of the magnetic field appears to induce a magnetic breakdown to the effect, and the system starts to behave more like a regular metal with a simple Fermi surface with no $\textbf{q}$-dependence. During this whole process, we should expect that the average of the interaction between conduction electrons and 4$f$ local moments is almost constant, which explains why there is an evolution of $\Delta g$ while $b$ remains unchanged. The last important piece of information to pinpoint our interpretation comes from the Gd$^{3+}$ spectra evolution as a function of the applied magnetic field. The transition from an insulating-like Gd$^{3+}$ spectrum at X-band to a collapse of the CEF splitting at Q-band data indicates a destructive-like interference of the carriers surrounding the Gd$^{3+}$ ions, which is the so-called local WAL effect.

A direct consequence of the local WAL effect, we may also understand the collapse of the CEF splitting of the Gd$^{3+}$ ions from a viewpoint of CEF contributions from the lattice and the carriers. The CEF parameters may be affected by the surrounding charges at the Gd$^{3+}$ site and eventually they can be strongly reduced. In other words, it appears that, due to the $S$-ion character of Gd$^{3+}$ ions, the influence of the conduction electrons could become relevant to the CEF effects compared to the lattice charges, as already shown in half-Heusler systems \cite{pagliuso1999crystal,souza2019crystalline}. As such, a change in the local charge distribution could cause a collapse of the Gd$^{3+}$ fine structure due to a strong reduction of the crystal field parameters associated with two contributions of different signs of the CEF parameters (lattice and carriers contributions). Since the second order crystal field parameter $b_{2}^{0}$ is much larger than the fourth order ones in (Bi$_{1-x}$Gd$_{x}$)$_{2}$Se$_{3}$ \cite{gratens1997epr}, we can estimate $|b_{2}^{0}|$ = 32/3*12 $\leq$ 0.9 Oe for Q-band measurements \cite{misra1986evaluation}.

One important concern about this proposed scenario is the comparison with previous transport results. Albeit previous reports show that even bulk Bi$_{2}$Se$_{3}$ samples have a WAL effect at low fields, and thinner samples have a more significant critical field ($H$ $\leq$ 1 T) \cite{kim2011thickness}, we must pay attention that all the results report WAL effects at low temperatures. In order to understand the origin of this discrepancy, it is important to understand that transport and ESR are different techniques. While the response in transport is a macroscopic, global property, in ESR we obtain a microscopic viewpoint. In ESR measurements we have two distinct relaxation channels: the relaxation through the spin-phonon process and the spin-spin relaxation, which is faster and involves the carriers of the system \cite{abragam2012electron,barnes1981theory,poole1971relaxation}. From our Gd$^{3+}$ spin dynamics, Fig. \ref{Fig3}, we clearly obtain signatures that the coupling between the 4$f$ local moments and the carriers are relevant, and they should dominate the relaxation of the system. Concomitantly, Gd$^{3+}$ has $L$ = 0, which means that the spin-phonon coupling is rather weak. That means that any potential scattering involving phonons, which would mask the WAL signatures, are going to be heavily suppressed. Such locality also help us to have an insight of the difference of critical fields. For thicker transport samples, the cusp has a critical field of only $\approx$ 1 kOe, while our X-band measurements occur at fields of 3.5 kOe. Again, the phonons contribution are going to be heavily suppressed, which means that it is understandable that the critical field for the local WAL effect might be comparable to the fields of thinner samples. Nonetheless, it is worth to note that Q-band measurements have fields applied at the order of $\approx$ 12 kOe. An alternate scenario could also rely on strain effects, which are originated from surface effects, playing a bigger role in Q-band measurements. In this scenario, the central line could have a smaller line width and we would be able to describe the data without an additional collapsed spectrum. Two different results indicate that this is an unlikely explanation. First of all, the data is better described with two resonances of different line widths at the same $g$-value. Even with we do not take this fact into account, the intensity of the central line should be, at least, double than the expected value to fairly describe the data - which is inconsistent with the crystal field Hamiltonian. Another point is that the skin depth is $\sim$ 11 $\mu$m and 8 $\mu$m for X- and Q-band measurements, respectively. With those skin depths, as expected, we do not see any signatures of surface effects in our data. Therefore, as far as our experimental data shows, the bulk dominates the ESR signatures and strain effects should not play a role.

It is worth also pointing the evolution of the sign of the $g$-shift. As already mentioned, our results indicate that $p$-type carriers are the main contributors near the Fermi level \cite{cao2013mapping,vidal2013photon}. However, earlier reports show positive $g$-shifts (typical for $s$-type carriers) for Gd$^{3+}$ substituted Bi$_{2}$(Te,Se)$_{3}$ \cite{isber1995hyperfine,gratens1997epr}. Perhaps, the higher temperature reached in their synthesis increases the number of defects, affecting the $g$-value - which is expected due to the Se and Te boiling points. Future systematic studies of Gd$^{3+}$ substituted Bi$_{2}$Se$_{3}$ synthesized at different temperatures could help to enlighten this open question.

The defects, mainly Se vacancies, also appears to have an important role in the interpretation of our experiments. First of all, we still observed a weak, but visible, Gd$^{3+}$ fine structure at Q-band measurements. Due to the different CEF states, Gd$^{3+}$ local moments closer to impurity centers may need higher fields to completely suppress the local WAL effect. Albeit it should be taken with care, it is possible to estimate the ratio between Gd$^{3+}$ ions in an insulating and conducting environment performing a double integration from our Gd$^{3+}$ ESR spectrum for Q-band measurements. We roughly estimate that 70 \% of the Gd$^{3+}$ ions have a conducting-like environment. As mentioned before, previous results indicate that vacancies tend to be inhomogeneous in Bi$_{2}$Se$_{3}$ matrix \cite{taylor2012spin}. Although the distribution of Gd$^{3+}$ ions appears to be homogeneous, it appears that those domains may have a tendency of accumulating near the magnetic impurities, which would be consistent with a high number of local moments center still showing an insulating character. The second hint of the role of vacancies comes from the Gd$^{3+}$ spin dynamics, especially the changes observed in the ESR data at low temperatures when comparing X-band and Q-band data. In this temperature range, presumably, the interaction between impurity centers and Gd$^{3+}$ local moments is more relevant, We obtained an evolution of the negative increase of the $g$-shift at low temperatures as a function of field: the Gd$^{3+}$ $g$-shift decrease due to spin-spin interactions is much more pronounced at X-band measurements when compared to Q-band ones [Fig. \ref{Fig3} c)]. The change in the charge distribution would change the interaction between defects and 4$f$ local moments and, naturally, we obtain a change in the low-$T$ Gd$^{3+}$ spin dynamics.

Regarding the topology of the system, although the strong spin-orbit coupling of the bulk is essential to the WAL mechanism, it is not clear if the topology of the system plays any role. Another potential signature of surface excitations could appear in a diffusive-like contribution to the Gd$^{3+}$ ESR line shape \cite{lesseux2016unusual,souza2018diffusive,souza2021surface}. However, the presence of carriers will naturally suppress any contribution from the surface excitations to the relaxation. This helps to understand, associated with the weak CEF effects of Gd$^{3+}$ ions, the lack of any diffusive-like effects in the Gd$^{3+}$ line shape \cite{lesseux2016unusual,souza2018diffusive,souza2021surface}. 

Going back to the magnetism of 4$f$ electrons in chalcogenides, the obtained exchange interactions, which is a Ruderman-Kittel-Kasuya-Yosida (RKKY) interaction, in Table \ref{TableI} are clearly smaller than those obtained for Bi$_{2-x}$Mn$_{x}$Te$_{3}$ \cite{zimmermann2016spin}. Additionally, it has been observed a change of the Korringa rate as a function of Mn$^{2+}$ concentration, which may be linked to changes in the $p-d$ hybridization. In the Gd$^{3+}$ case, we observed the same Korringa relaxation for all the explored concentrations, indicating that $f-p$ hybridization may only play a small role, highlighting the clear difference between the effects of $d$ and $f$ substitutions. 

\section{\label{sec:conclusion}V. Conclusion}

In summary, we performed electron spin resonance and complementary macroscopic measurements in the Gd$^{3+}$-substituted Bi$_{2}$Se$_{3}$ ($x$ = 0, 0.001, 0.002 and 0.006). The Gd$^{3+}$ ESR spectra at 4 K for different concentrations show seven Dysonian lines for X-band measurements, which evolves to an apparent contribution from distinct Gd$^{3+}$ sites spectra for Q-band. We conjecture that such evolution might be due to a breakdown of the local WAL effect and a change in the crystal field parameters in the vicinity of the vacancies. This interpretation is consistent with the Gd$^{3+}$ spin dynamics response. Additionally, we show that the 4$f$ substitution does not increase the DOS at the Fermi level, neither introduces a relevant $f-p$ hybridization. This is in contrast to more traditional substitutions with transition metals $d$ states magnetic ions and indicates different magnetic mechanisms for $d$ and $f$ states in this system. Our work points out that 4$f$ substitution in chalcogenides is an interesting path to explore even further the role of magnetic impurities in this model system.

\begin{acknowledgments}

We would like to thank Dr. S. K. Misra for providing Ref. \cite{misra1986evaluation}. This work was supported by FAPESP\ (SP-Brazil) Grants No 2022/09240-3, 2020/12283-0, 2019/26247-9, 2018/11364-7, 2017/10581-1, 2012/04870-7, CNPq Grants No 311783/2021-0, 309483/2018-2, 442230/2014-1 and 304649/2013-9, CAPES and FINEP-Brazil. Work at Los Alamos was supported by the Los Alamos Laboratory Directed Research and Development program through project 20210064DR.

\end{acknowledgments}


%


\end{document}